\documentclass[11pt,twoside]{article}

\usepackage{asp2006}
\usepackage{epsf}
\usepackage{psfig}
\usepackage{lscape}

\begin{document}
\title{Polarization from the Structured Envelopes of Cool Evolved Stars}   
\author{R.~Ignace$^1$, G.~D.~Henson$^1$, J.~Carson$^2$}   
\affil{
$^1$ Department of Physics, Astronomy, \& Geology, East Tennessee State
University \\
$^2$ Department of Physics \& Astronomy, Western Kentucky University}    

\begin{abstract} 
We present preliminary calculations of electron scattering polarizations
from models of structured cool star envelopes.  We note that net
polarizations from unresolved sources can result from non-spherical
scattering envelopes and/or anisotropic illumination from a photosphere
that has brightness variations.  The resultant polarizations are
quite small (hundreths of a percent); however, Rayleigh scattering
from molecular opacity and/or dust scattering from the more extended
envelope under similar considerations may produce higher polarizations.
\end{abstract}

\begin{figure}
\epsfxsize=.8\textwidth
\plottwo{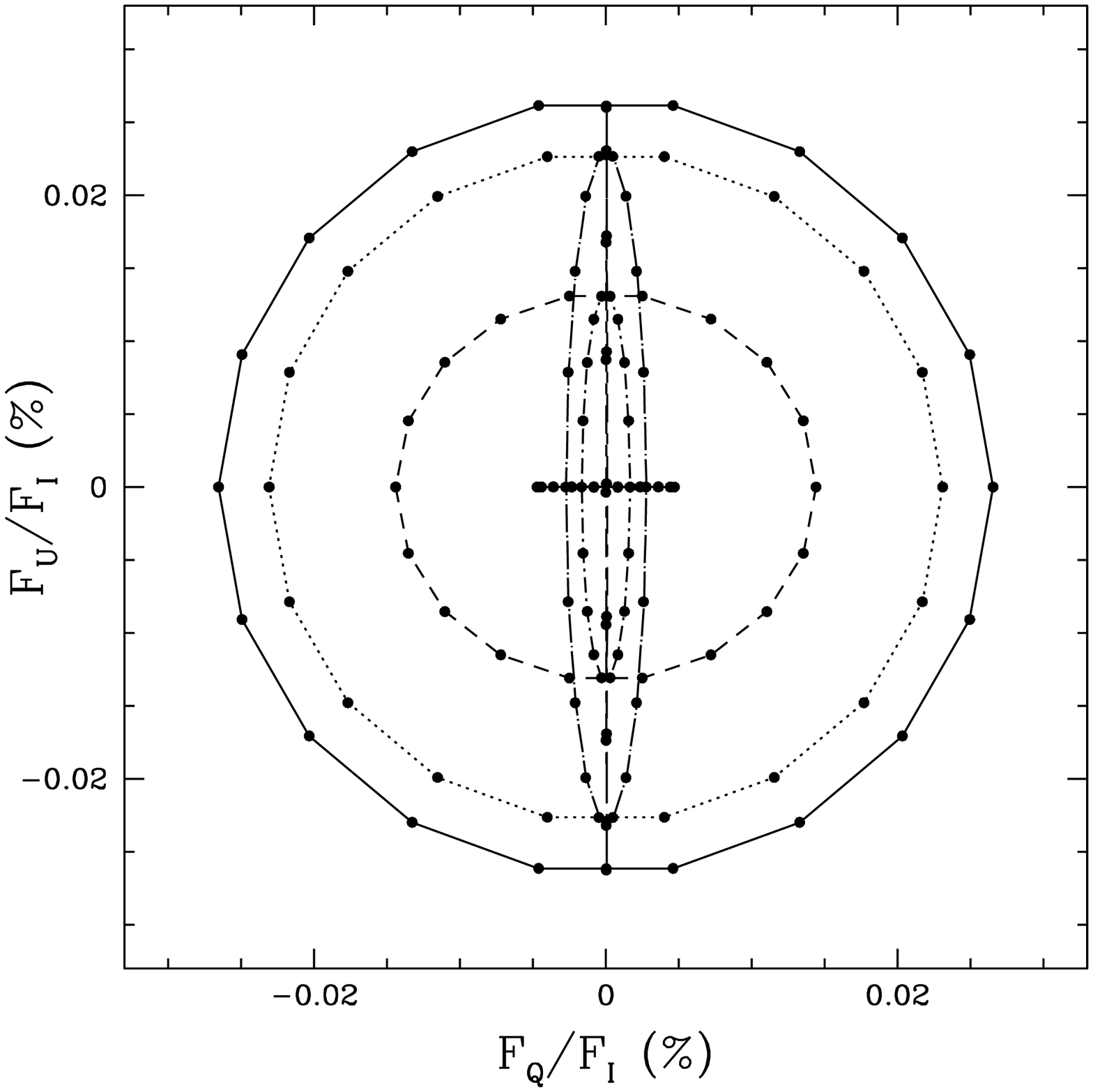}{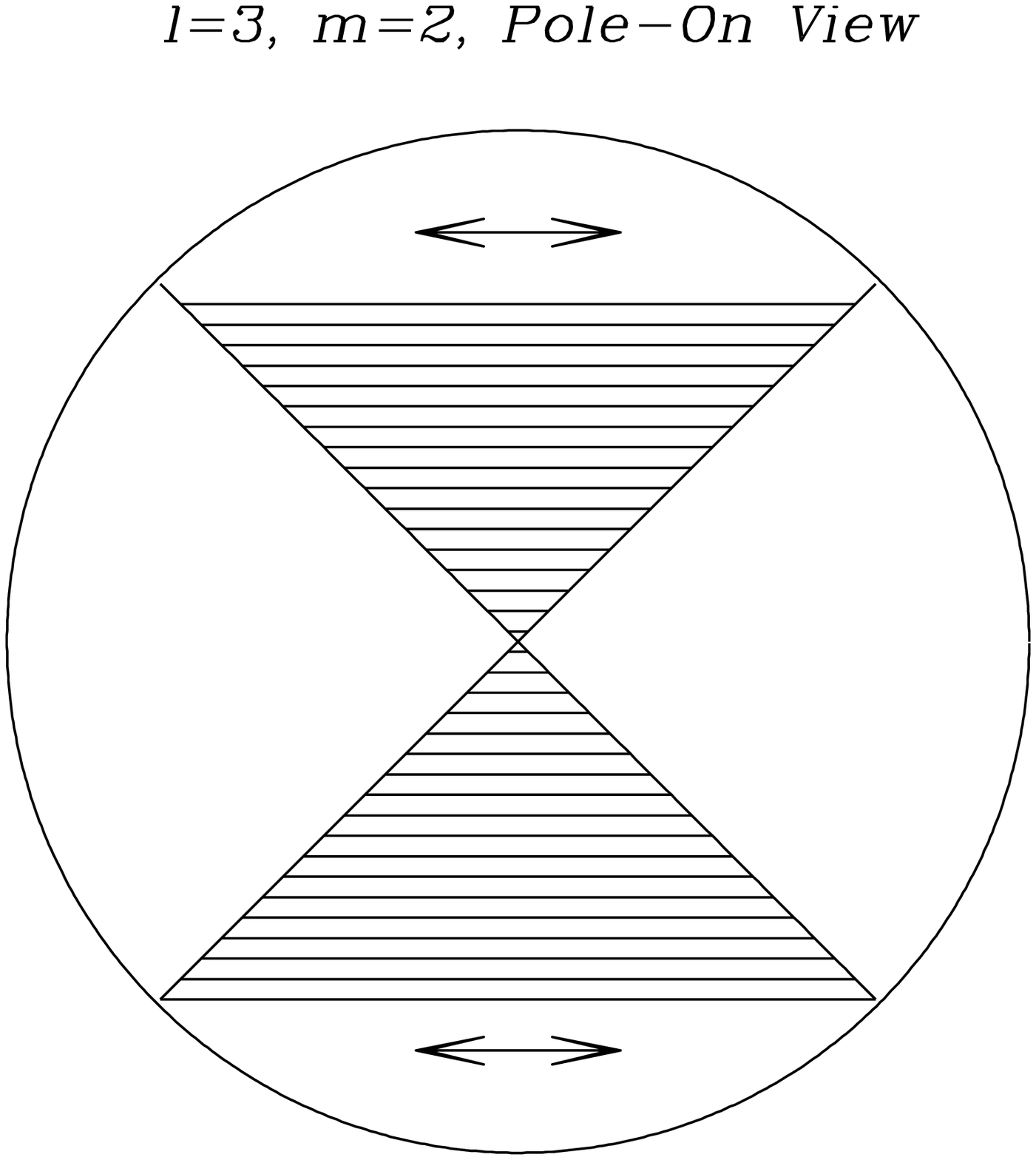}
\caption{{\it Left:} Polarization from a structured chromosphere as seen from
different viewing inclinations (curves) and azimuthal orientations
(points). {\it Right:}  Representation of the pole-on appearance for
the $l=3, m=2$ density distribution.  The lined wedges are higher density
with associated polarization position angle given by the arrows.
Rotating the figure results in the large circle of the $Q-U$ diagram
at left.
\label{ignace:fig1}}
\end{figure}

The variable nature and complex envelopes of cool evolved stars offer
numerous mechanisms for the creation of polarized light, which in turn
probes structure in the envelopes and flows of these star.  Observations
of variable polarization lead to constraints on physical models for the
envelope dynamics.  We consider an electron scattering chromosphere
with a radial density of scatterers $n \propto \exp[(r-R)/H)]
\times \sin^2[k(r-R)]$ illuminated by a photosphere at $r=R$.
Figure~\ref{ignace:fig1} shows a $Q-U$ diagram for a chromospheric
envelope with $\tau=1$, scale height $H=0.01R$, and lateral structure
given by a spherical harmonic of $l=3, m=2$.  Different curves are
for different viewing inclinations, and the points are for different
azimuthal perspectives.  The large circle is for a pole-on view, with
progressive curves increasing the viewing angle by $15^\circ$ with 
the vertical line representing the equatorial view.

\begin{figure}
\plottwo{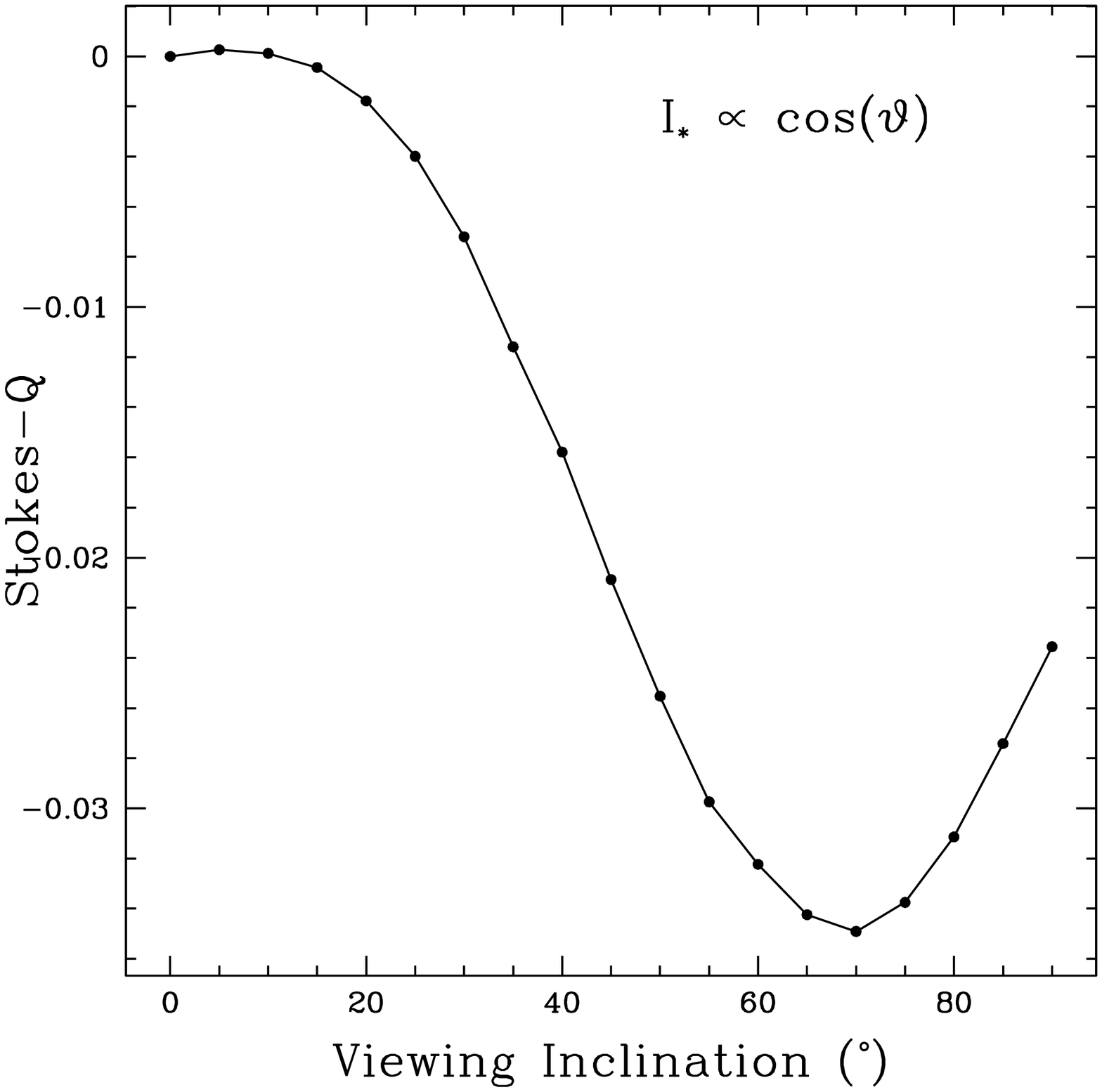}{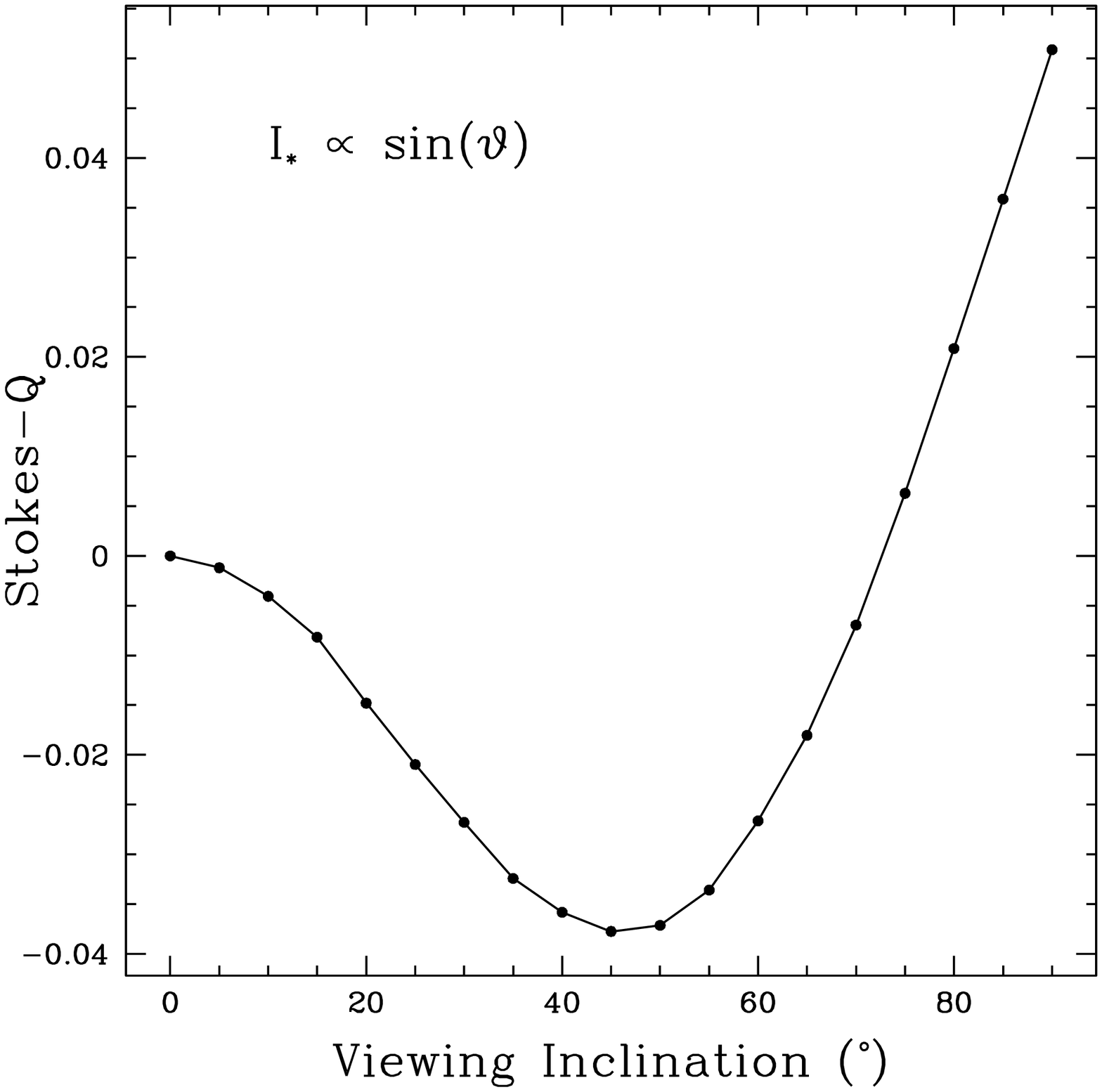}
\caption{A spherically symmetric envelope illuminated by a polar-bright
(left) or equatorial-bright (right) star.  Here $\vartheta$
is the co-latitude, and the polarizations are percent values.
\label{ignace:fig2}}
\end{figure}

Expanding on Al-Malki (1992)
and Al-Malki et~al.\ (1999), we consider the converse case
of a spherically symmetric (but radially structured) chromosphere
that is anisotropically illuminated.  Results are shown in
Figure~\ref{ignace:fig2}:  left for a pole-bright star, and right
for an equator-bright star.  Note that these are axisymmetric. 

\acknowledgements

This project was partially funded by a partnership between the NSF
(AST-0552798), the REU, and the DoD ASSURE programs.

\end{document}